\documentclass[twocolumn,tighten,times]{aastex63}

\usepackage{amsmath}
\usepackage[caption=false]{subfig}
\usepackage{empheq}

\newcommand{\Mtot}{$M_{\mathrm{bin}}$}

\newcommand{\fgd}{$f_{\mathrm{gd}}$}
\newcommand{\vg}{$v_{\rm g}$}

\newcommand{\rhog}{$n_{\rm gd0}$}
\newcommand{\tevol}{$t_{\mathrm{evol}}$}

\shorttitle{The Pairing of MBHs in the Presence of Radiative Feedback}
\shortauthors{Li et al.}

\begin{document}

\title{The Pairing Probability of Massive Black Holes in Merger
  Galaxies in the Presence of Radiative Feedback}

\author[0000-0002-0867-8946]{Kunyang Li}
\affiliation{School of Physics and Center for Relativistic
  Astrophysics, 837 State St NW, Georgia Institute of Technology,
  Atlanta, GA 30332, USA}   
\email{kli356@gatech.edu}

\author[0000-0002-7835-7814]{Tamara Bogdanovi{\'c}}
\affiliation{School of Physics and Center for Relativistic
  Astrophysics, 837 State St NW, Georgia Institute of Technology,
  Atlanta, GA 30332, USA}
\email{tamarab@gatech.edu}

\author[0000-0001-8128-6976]{David R. Ballantyne}
\affiliation{School of Physics and Center for Relativistic
  Astrophysics, 837 State St NW, Georgia Institute of Technology,
  Atlanta, GA 30332, USA}
\email{david.ballantyne@physics.gatech.edu}




\begin{abstract}
Dynamical friction (DF) against stars and gas is thought to be an important mechanism for orbital evolution of massive black holes (MBHs) in merger remnant galaxies. Recent theoretical investigations however show that DF does not always lead to MBH inspiral. For MBHs evolving in gas-rich backgrounds, the ionizing radiation that emerges from the innermost parts of their accretion flow can 
affect the surrounding gas in such a way to cause the MBHs to
accelerate and gain orbital energy. This effect was dubbed ``negative
DF". We use a semi-analytic model to study the impact of negative DF
on pairs of MBHs in merger remnant galaxies evolving under the
combined influence of stellar and gaseous DF. Our results show that
for a wide range of merger galaxy and MBH properties negative DF
reduces the MBH pairing probability by $\sim 46\%$. The suppression of MBH pairing is most severe in galaxies with one or more of these properties: (1) a gas fraction of $f_g \geq 0.1$; (2) a galactic gas disk rotating close to the circular velocity; (3) MBH pairs in prograde, low eccentricity orbits, and (4) MBH pairs with mass $< 10^8\,$M$_\odot$. The last point is of importance because MBH pairs in this mass range are direct progenitors of merging binaries targeted by the future space-based gravitational wave observatory LISA.
\end{abstract}


\keywords{Dynamical friction (442) --- Galaxy dynamics (591) --- Galaxy evolution (594) --- Galaxy mergers (608) --- Supermassive black holes (1663)}




\section{Introduction}
\label{sec:intro}

Massive black holes (MBHs), with masses in the range $\sim 10^6 -
10^{10}\, {\rm M}_{\odot}$, are known to exist in the centers of most
galaxies \citep{S1982, KR1995, M1998}. After two galaxies merge, a MBH
pair\footnote{We refer to the system of two
  MBHs as a MBH pair when they are not gravitationally bound, and as a
  MBH binary (MBHB) when they are gravitationally bound.} may inspiral
in the remnant galaxy and coalesce due to the emission of
gravitational waves \citep[GWs;][]{BBR1980}. At separations of $\sim
1$ kpc, the orbital decay of MBHs is expected to be driven by
dynamical friction (DF) by stars and gas \citep{C1943,O1999} in the
remnant galaxy. The evolution timescale of such pairs to separations where they form gravitationally bound binaries is determined by the properties of the two MBHs and their host galaxy. 

In earlier work \citep[][hereafter LBB20]{LBB2020}, we found that the
percentage of MBHs that form gravitationally bound binaries within a
Hubble time is $> 80\%$ in remnant galaxies with gas fractions $< 20\%$, and in galaxies hosting MBH pairs with total mass $ > 10^6$ M$_{\rm \odot}$ and mass ratios $\gtrsim1/4$. Among these, the remnant galaxies with the fastest formation of MBHBs have at least one of these properties: large stellar bulge, comparable mass MBHs, and a galactic gas disk rotating close to the circular velocity. In such galaxies, the MBHs with the shortest inspiral times are either on circular prograde orbits or on very eccentric retrograde orbits. These MBHs are the most likely progenitors of coalescing binaries, whose GWs are expected to be detected by the pulsar timing arrays \citep[PTAs;][]{PTA1990, shannon2015, Lentati2015, Arz2018} and the Laser Interferometer Space Antenna \citep[LISA;][]{LISA2017, Klein2016} in the next few to 15 years. 

\begin{deluxetable*}{ccC}
\tablenum{1}
\tablecaption{Galaxy Model Parameters\label{tab:params}}
\tablewidth{0pt}
\tablehead{
\colhead{Symbol} & \colhead{Description} & \colhead{Values}
}
\startdata
\Mtot\ & total MBH pair mass & (2,3,5)\times 10^5 \mathrm{M}_{\odot}\\
 	& & (1,3)\times 10^6 \mathrm{M}_{\odot}\\
 	 & & (1,3)\times 10^7 \mathrm{M}_{\odot}\\
 	 & & (1,3)\times 10^8 \mathrm{M}_{\odot}\\
 $q$ & MBH mass ratio & 1/n\,\,  (n=2,\ldots,9) \\
 \rhog\ & central gas number density & 100, 200, 300\,{\rm cm}$^{-3}$ \\
 \fgd\ &  gas disk mass fraction & 0.3, 0.5, 0.9 \\
$v_{\rm g}(r)$ \ & gas disk rotational speed in steps of $0.1 v_{\rm c}(r)$
 & -0.9$v_{\rm c}(r)$,\ldots,0.9$v_{\rm c}(r)$ \\
\enddata
\tablecomments { $v_{\rm g} > 0$ ($v_{\rm g} < 0$) corresponds to the sMBH corrotating (counterrotating) with the galactic disk.}
\end{deluxetable*}

Inspiral and coalescence of MBH pairs within a Hubble time
is not a inevitable, even when merger galaxies and their MBHs
have the properties described above. For example, it was recently
shown for MBHs evolving in gas-rich backgrounds that ionizing
radiation emerging from the innermost parts of the MBHs' accretion
flows can affect their gaseous DF wake and render gas DF
inefficient for a range of physical scenarios. MBHs in this regime
tend to experience positive net force, meaning that they speed up,
contrary to the expectations for gaseous DF without radiative feedback \citep{PB2017,G2020,T2020}. This effect, dubbed ``negative DF", is only present when the system satisfies the following criteria \citep{I2016, PB2017}:
\begin{eqnarray}
\label{eq:cond1}
(1+{\cal M}^2)M_{\rm bh}n_{\rm \infty}  <10^9{\rm M}_{\rm \odot}\,{\rm cm}^{-3}\, T_{4}^{1.5} ,\; {\cal M}< 4\,
\end{eqnarray}
where $M_{\rm bh}$ is the mass of the orbiting MBH, $n_{\rm \infty}$ is the gas number density ``at infinity", unaffected by the gravity of the MBH, and $T_{4} = T/10^4\,{\rm K}$ is the gas temperature at the position of the orbiting MBH. Here, ${\cal M} = \Delta v/c_{\rm s}$ is the Mach number, $ \Delta v$ is the speed of the MBH relative to the gas, $c_s = \sqrt{5kT/3m_{\rm p}}$ is the sound speed, and other constants have their usual meaning. 

The first criterion in equation~\ref{eq:cond1} provides a limit within
which the size of the ionized region around the MBH is larger than its trailing gaseous DF wake, and so the ionizing radiation
suppresses its formation. Without the dense trailing wake, the MBH is
pulled ``forward" and accelerated by the dense shell of gas that forms
in front of the MBH due to the ``snowplow" effect caused by radiation pressure \citep{PB2017}. 

However, MBHs under the influence of negative
DF do not perpetually accelerate. According to the second criterion,
which follows directly from the jump conditions for ionization fronts
\citep{Park2013}, a limit exists for the maximum MBH
velocity that can be achieved due to negative DF. This limit suggests
that, without any other external forces, MBHs that are
subject to negative DF should move with an equilibrium ${\cal M} \sim \mathrm{few}$. MBHs that do not satisfy these criteria, either because their speed corresponds to ${\cal M} \gtrsim 4$, or because they are embedded in sufficiently high density gas, are subject to classical gaseous DF described by \citet{O1999}.

If prevalent in real merger galaxies, negative gaseous DF can lengthen
the inspiral time of MBHs. Its implications for the formation and
coalescence rate of MBHBs in galactic and cosmological settings are
however yet to be understood. Our work is an extension of earlier
studies that employed N-body simulations of MBH pairs in stellar
environments \citep[][and others]{Q1996, QH1997, Y2002, B2006,KJM2011,
  K2013}, hydrodynamic simulations of MBH pairs in gas-rich
environments \citep{E2005, D2007, C2009}, and semi-analytic models of
MBH orbital decay \citep{AM2012, B2012, K2016, B2016, DA2017,
  KBH2017,K2017b}. Here, we consider the DF from both stars and gas in
the galaxy, and, for the first time, quantify the effect of negative gaseous DF on the inspiral time and pairing probability of MBHBs.

\section{Methods}
\label{sec:methods}

In this work we build upon a semi-analytic model presented by LBB20,
which describes the orbital evolution of MBH pairs under the influence
of stellar and gaseous DF without radiative feedback. In this model we assume that a single remnant galaxy, that forms after the galaxy merger, hosts the MBH pair.  The remnant consists of a stellar bulge, stellar disk, and a gas disk. 

The radial density of the stellar bulge is described using a power-law
profile, truncated at the characteristic outer radius \citep{BT1987},
and has a total mass proportional to the primary MBH (pMBH)
\citep{M1998}, with the proportionality constant equal to 1000. The
stellar and gas disks both follow exponential density profiles
\citep{BT1987}. The temperature profile for the gas disk is calculated
using the Toomre stability criterion \citep{T1964}, which gives the
minimum temperature for which the gas disk is stable to gravitational
collapse. We set the temperature of the disk to be $10^4$\,K above
this minimum temperature, since the interstellar medium after a
galactic merger is likely to be shocked and turbulent \citep[e.g.,][]{BH1991}. Therefore, this gas temperature should be interpreted as a proxy for unmodeled turbulence of warm gas.

\begin{figure*}[t]   
    \centering
        \includegraphics[width=0.49\textwidth]{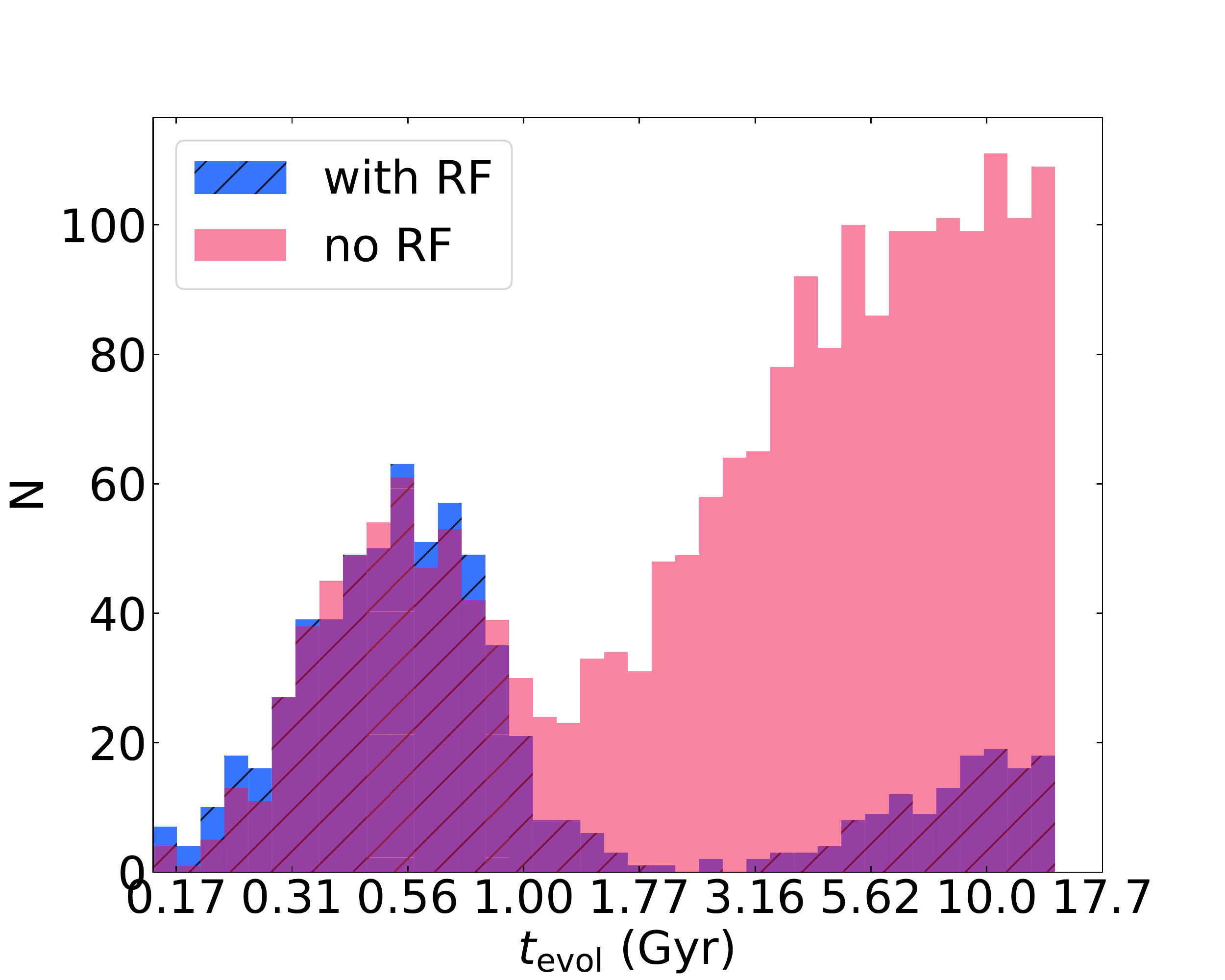}
        \includegraphics[width=0.49\textwidth]{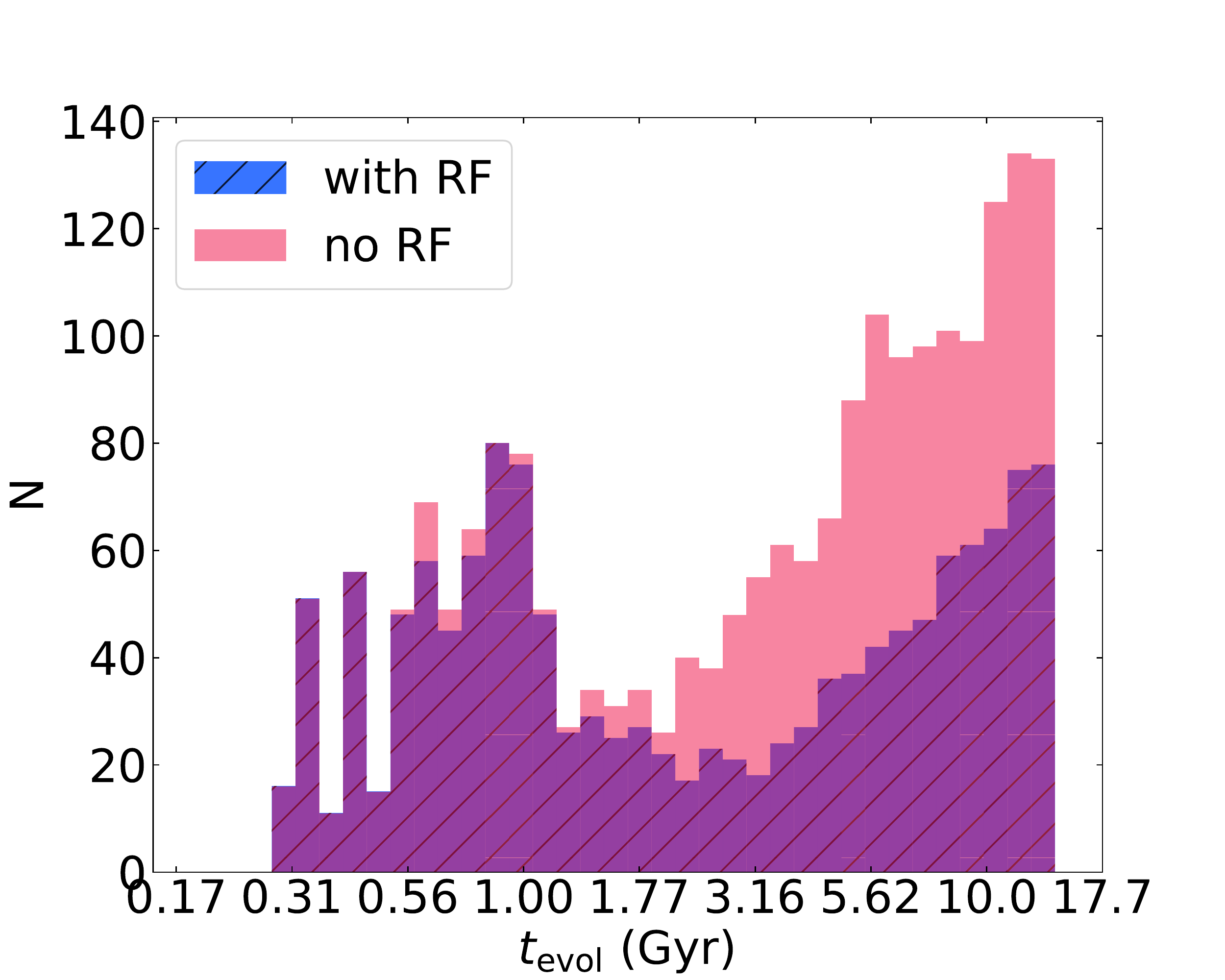}
   \caption{{\it Left:} Histograms of inspiral time for sMBHs on
     prograde, low $e_{\rm i}$ ($e_{\rm i} < 0.2$) orbits in simulations with and without
     radiative feedback. {\it Right:} As in the left-hand panel, but
     now for sMBHs on retrograde, low $e_{\rm i}$ orbits. The $y$-axis
     in both panels shows the number of MBH pairs with \tevol\ shorter than a Hubble time.}
\label{time_hist}
\end{figure*}

Each remnant galaxy is described by five parameters summarized in
Table~\ref{tab:params}. Parameter $M_{\rm bin} =  M_1 + M_2$ is the
total mass of the MBH pair and $q = M_2/M_1 < 1$ is the mass ratio of
the secondary MBH (sMBH) to the pMBH. \rhog\ is the number density of
gas particles at the center of the remnant galaxy. Parameter \fgd\ is
the gas disk mass fraction defined as $f_{\rm gd}= M_{\rm
  gd,1}/(M_{\rm gd,1}+M_{\rm sd,1})$, where $M_{\rm gd,1}$ and
$M_{\mathrm{sd,1}}$ are the masses of the gas and stellar disks within
1\,kpc, respectively. We assume that the gas and stellar disk are
rotating with the same speed, $v_{\mathrm{g}}(r)$, while the bulge
does not rotate. The parameter $v_{\mathrm{g}}(r)$ is in units of circular velocity $v_{\mathrm{c}}(r)$ and $v_{\rm g} > 0$ indicates that the MBH pair corrotates with the disk and vice versa. The parameter grid shown in Table~\ref{tab:params} corresponds to 39366 model galaxies. 

We focus on the evolution of unequal mass MBH pairs and fix the location
of the pMBH to the center of the remnant. The sMBH orbits the center
of the galaxy (and the pMBH) on an orbit that is always coplanar  with
the galactic disk. We also assume that the sMBH is stripped of its
nuclear star cluster in the early stages of the galactic merger preceding the starting point of our simulations. The sMBH is subject to stellar DF exerted by the bulge and stellar disk, and gaseous DF due to the gas disk. We calculate the stellar DF force following equations~(5)-(7) in LBB20, based on calculations presented by \citet{AM2012}. The velocity distribution of stars in the bulge is set to be Maxwellian:
\begin{equation}
\label{eq:maxw}
f{\rm (v_\star)} = \frac{1}{(2\pi\sigma^2_\star)^{3/2}}e^{-v^2_\star/2\sigma^2_\star},
\end{equation}
where $\sigma_\star$ is the velocity dispersion of bulge stars, estimated from the $M - \sigma$ relation for the primary MBH \citep{G2009, MM2013}.

To evaluate the DF force due to gas, we first use the criteria in equation~\ref{eq:cond1}, to determine whether the sMBH is in the regime where gas DF is affected by radiative feedback. If the criteria are not fulfilled, we calculate the gaseous DF force following equations~10--12 in LBB20, corresponding to the case ``no RF" below. When the criteria are fulfilled, we use the modified expression for the gaseous DF force shown in the case ``with RF".
\begin{equation}
\label{eq:gdforce}
\vec{F}_{\rm gd} = -\frac{4\pi (GM_2)^2 \rho_{\rm gd}}{\Delta v^2}
  	\begin{cases}
        	I_R \hat{R} + I_{\phi} \hat{\phi} \;\;\;\;\;\;\text{(no RF)},\\
        	I_R \hat{R} - 0.6 I_{\phi} \hat{\phi} \;\;\text{(with RF)}.
        	\end{cases}
\end{equation}
Here, $\Delta v$ is the velocity of the sMBH relative to the gas disk
and $\rho_{\rm gd}$ is the gas density defined by equation~3 of
LBB20. $I_R$ and $I_{\phi}$ are the dimensionless components of the DF
force in the radial and azimuthal directions, defined by \citet{KK2007} and adopted by LBB20. Both $I_R$ and $I_{\phi}$ are functions of the Mach number and peak sharply when ${\cal M} = 1$. Since commonly $I_{\phi}\gg I_R$, the gaseous DF force is strongest when $\Delta v$ is close to the sound speed of the gas. 

The key modification in the ``with RF" case of equation~\ref{eq:gdforce} is motivated by the finding that for a MBH moving through gas on a linear trajectory, the magnitude of the negative DF force is $\sim 60$\% of that expected without radiative feedback \citep{PB2017}. We neglect the effect of radiative feedback on the radial component of the DF force, which does not impact the results significantly since $I_{\phi}\gg I_R$.

Over the course of each simulation, we record the farthest and closest radial distance of the sMBH from the pMBH for every orbit and use them to estimate the orbital semimajor axis ($a$) and eccentricity ($e$). Since the galaxy potential is not proportional to $1/r$, and the orbits described by the sMBH are neither Keplerian nor closed, the computed values of $a$ and $e$ are only used to illustrate the shape and size of the orbits. In those terms, the sMBH in each simulation starts on an orbit with $a_{\rm i} \sim 1$ \,kpc and initial eccentricity, $e_{\rm i}$. The simulations are stopped when the sMBH reaches a separation of 1\,pc from the center for the first time.


\section{Effect of Radiative Feedback on the Pairing Probability of MBHs}
\label{sec:results}

Below, we investigate the effect of negative gas DF on the
inspiral time (\tevol) and pairing probability of MBHs in merger
galaxies with different properties. Figure~\ref{time_hist} shows
histograms of \tevol\ in the scenario where we either account for or
neglect radiative feedback. In all cases we calculate \tevol\ as the
time for the MBH pair to evolve from an initial separation of $\sim
1$\,kpc down to 1\,pc and the plots show only systems that complete the evolution in less than a Hubble time. 

\begin{figure*}[t]   
    \centering
        \includegraphics[width=0.8\textwidth]{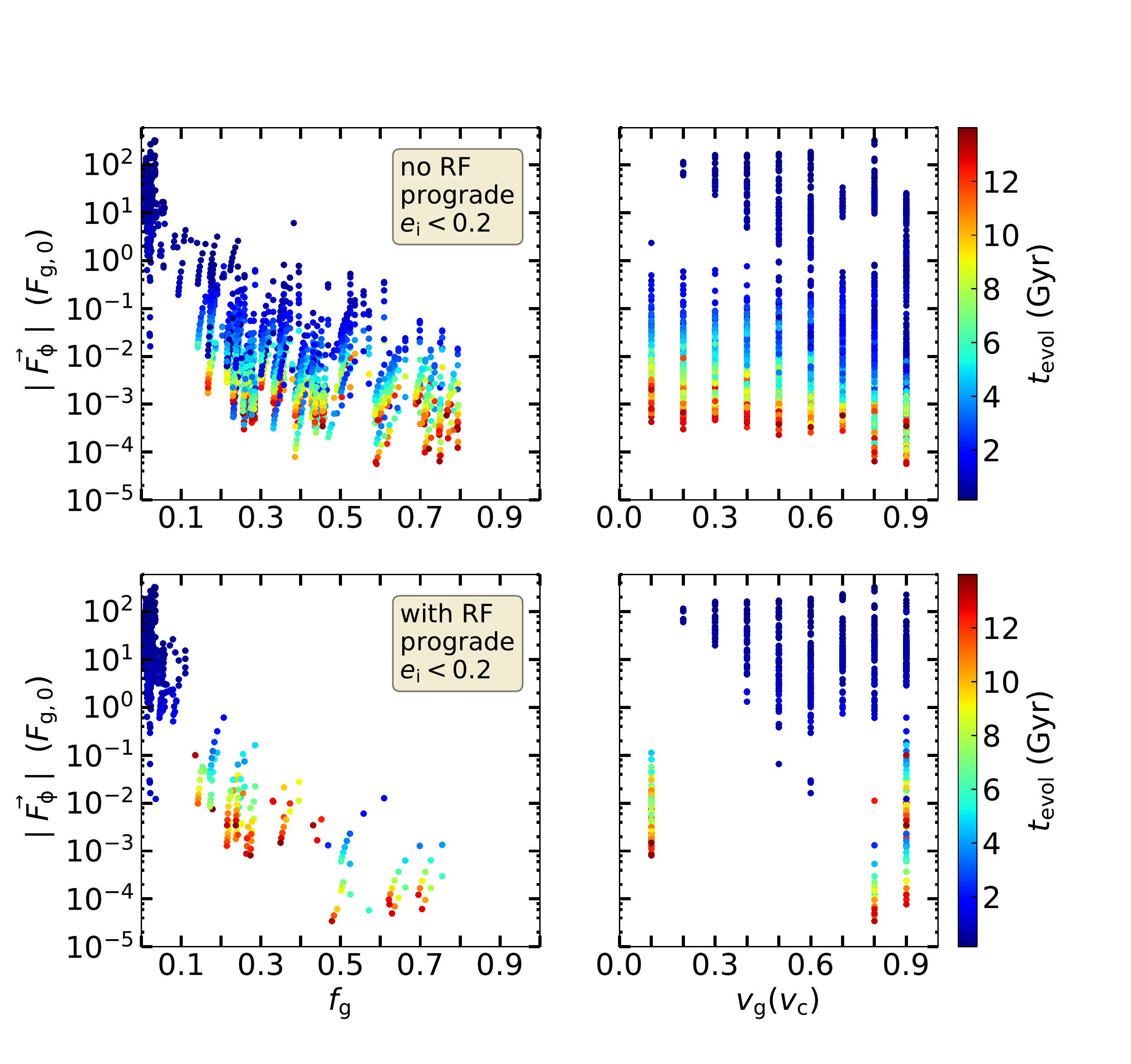}
 \caption{The relationship between the two parameters of the model ($f_{\rm g}$ and \vg), the total DF force, and the inspiral time (\tevol), for MBH pairs with prograde orbits and low initial orbital eccentricity. We show the time averaged azimuthal component of the force, $\vec{F}_{\phi}$, that is responsible for the orbital evolution and neglect the radial component. The top (bottom) row of panels corresponds to simulations without (with) radiative feedback. The color marks the inspiral time. The MBH pairs with inspiral time longer than a Hubble time are not shown in this figure. }
\label{prop_l}
\end{figure*}

Figure~\ref{time_hist}(left) illustrates the results for
MBHs on prograde orbits with low $e_{\rm i}$ ($e_{\rm i} <
0.2$). Without radiation feedback these configurations are
characterized by a bimodal distribution of \tevol. The left peak of the histogram (at $\sim 0.56$\,Gyr)
corresponds to galaxies in which the stellar bulge dominates the
orbital evolution of the sMBH, while gaseous DF dominates the evolution
for models in the right peak 
(at $\sim 10$\,Gyr). The difference in the two populations arises because of a significantly larger magnitude of the DF force exerted by the bulge (see \S~3.3 in LBB20 for discussion).

\begin{figure*}   
    \centering
        \includegraphics[width=0.8\textwidth]{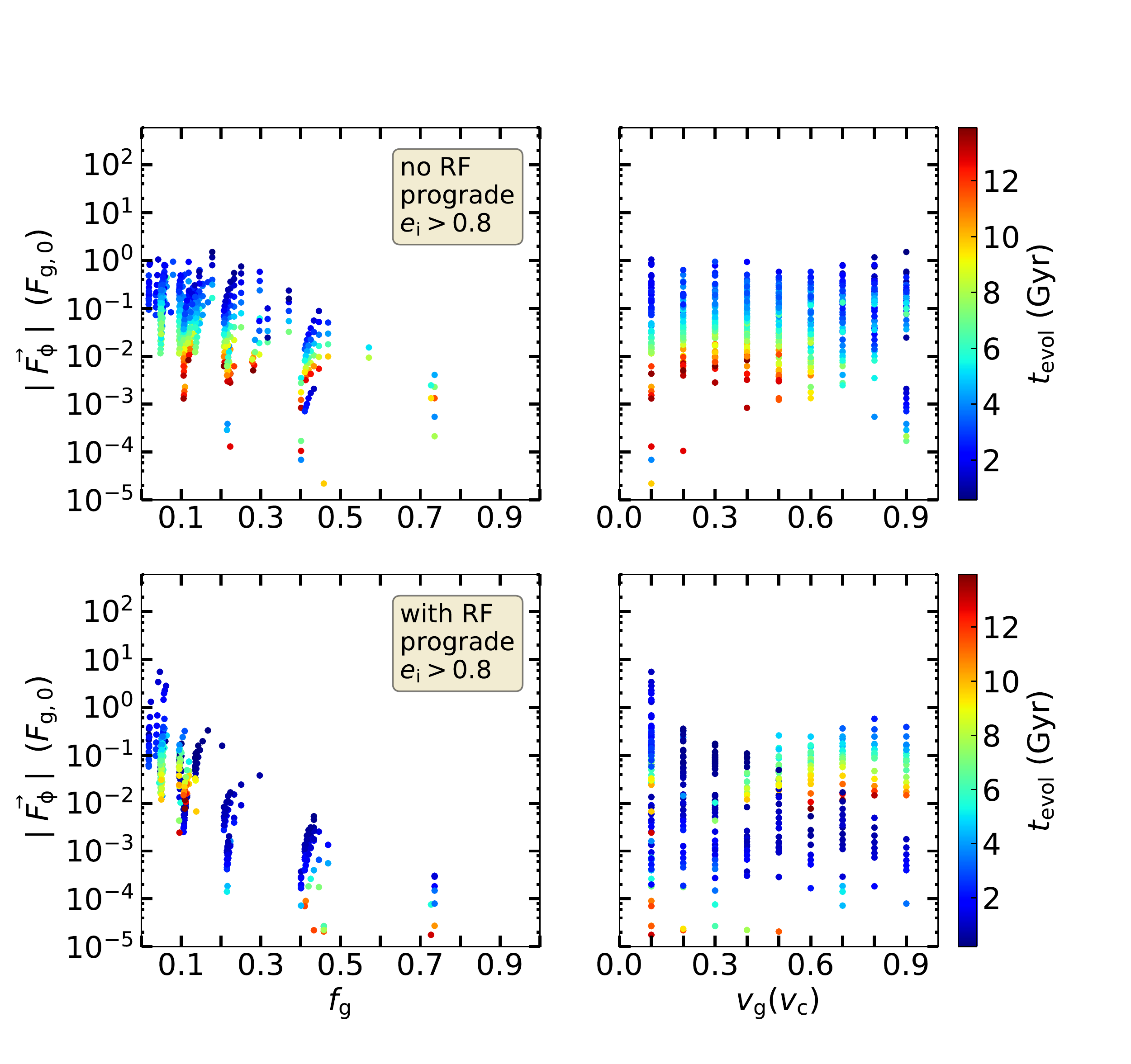}
   \caption{Same as Figure~\ref{prop_l} but for MBH pairs in prograde orbits with high initial orbital eccentricity. }
\label{prop_h}
\end{figure*}

In the presence of radiative feedback the distribution of
\tevol\ remains bimodal but its right peak is significantly reduced. In comparison, \tevol\ of the MBH pairs whose evolution
is dominated by the stellar bulge (the left peak) is only weakly
affected. This is because negative DF is more pronounced for MBHs
whose orbital evolution is determined by gas. Overall, the number
  of MBH pairs in prograde, low $e_{\rm i}$ orbits that reach 1\,pc
  within a Hubble time is reduced from 2170 without
  radiative feedback to 702 with radiative feedback (a reduction of 67\%).

Figure~\ref{time_hist}(right) shows \tevol\ for sMBHs on
retrograde orbits with low $e_{\rm i}$. The distribution of
\tevol\  is again bimodal, with the left peak corresponding to MBH
pairs that evolve due to DF largely exerted by the bulge, and the
right peak corresponding to pairs whose evolution is dominated by
gaseous DF. For retrograde orbits too, the difference between the two
histograms is largest for systems evolving as a consequence of the gas
torques. In this case, however, the right peak is not suppressed as
severely in the presence of radiative feedback as for the prograde orbits. This is because the sMBHs in retrograde orbits have larger velocities relative to the gas disk, resulting in ${\cal M}>4$ and restored classical gas DF. Consequently, negative DF reduces the number of MBH pairs in retrograde, low $e_{\rm i}$ orbits that reach 1\,pc within a Hubble time from 2083 to 1364 (a reduction of 35\%). 

We evaluate the magnitude of the DF force for sMBHs
on prograde orbits (since, as seen above, these tend to be
more affected by negative DF) as a function of two key model
parameters: the gas fraction, $f_g$, and \vg. Here, $f_g = M_{\rm
  gd}/(M_{\rm gd}+M_\star)$ is a parameter that can be compared
directly with the gas fraction of galaxies inferred from observations
and  $M_{\mathrm{\star}}$ includes both the mass of the bulge and
stellar disk within the central kiloparsec. LBB20 found that, without
radiation feedback, the orbital evolution of MBH pairs in galaxies
with $f_{\rm g} < 0.2$ tends to be
dominated by stellar bulges, and in those with $f_{\rm g} \geq 0.2$ it
is dominated by classical gaseous DF.

\begin{figure*}[t]
\centering
        \begin{tabular}{@{}cc@{}}
            \includegraphics[width=0.49\textwidth]{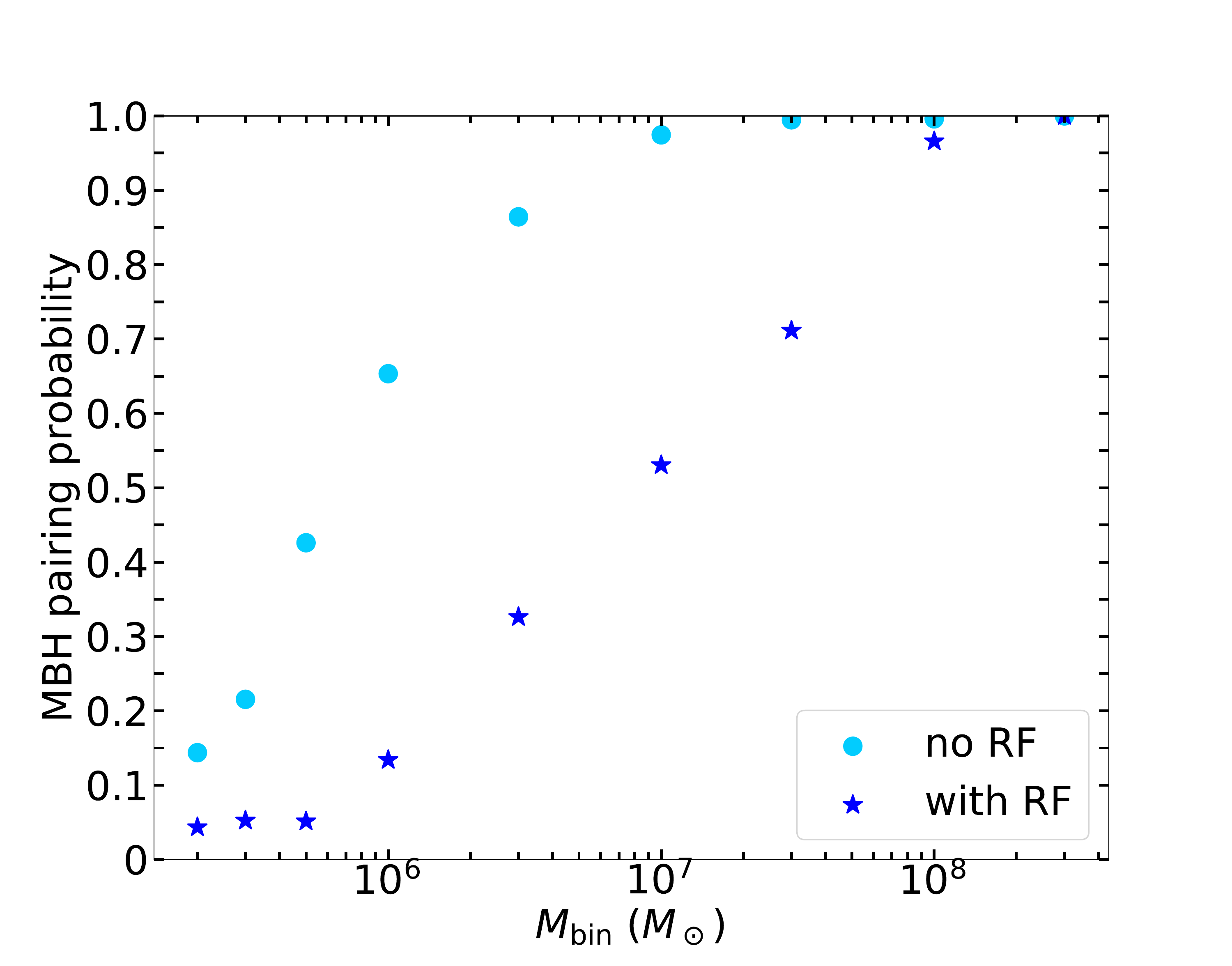}
            \includegraphics[width=0.49\textwidth]{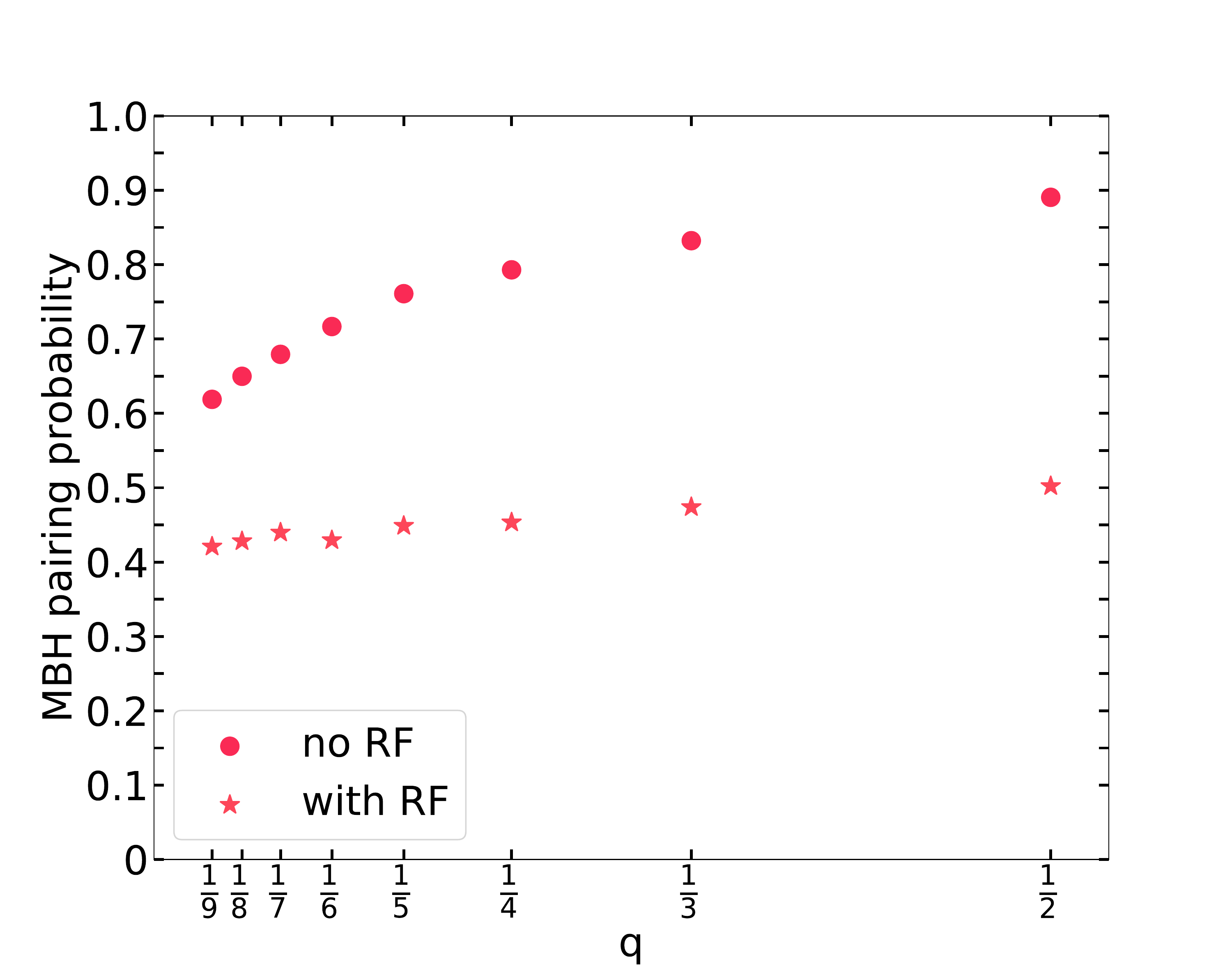}\\
            \includegraphics[width=0.49\textwidth]{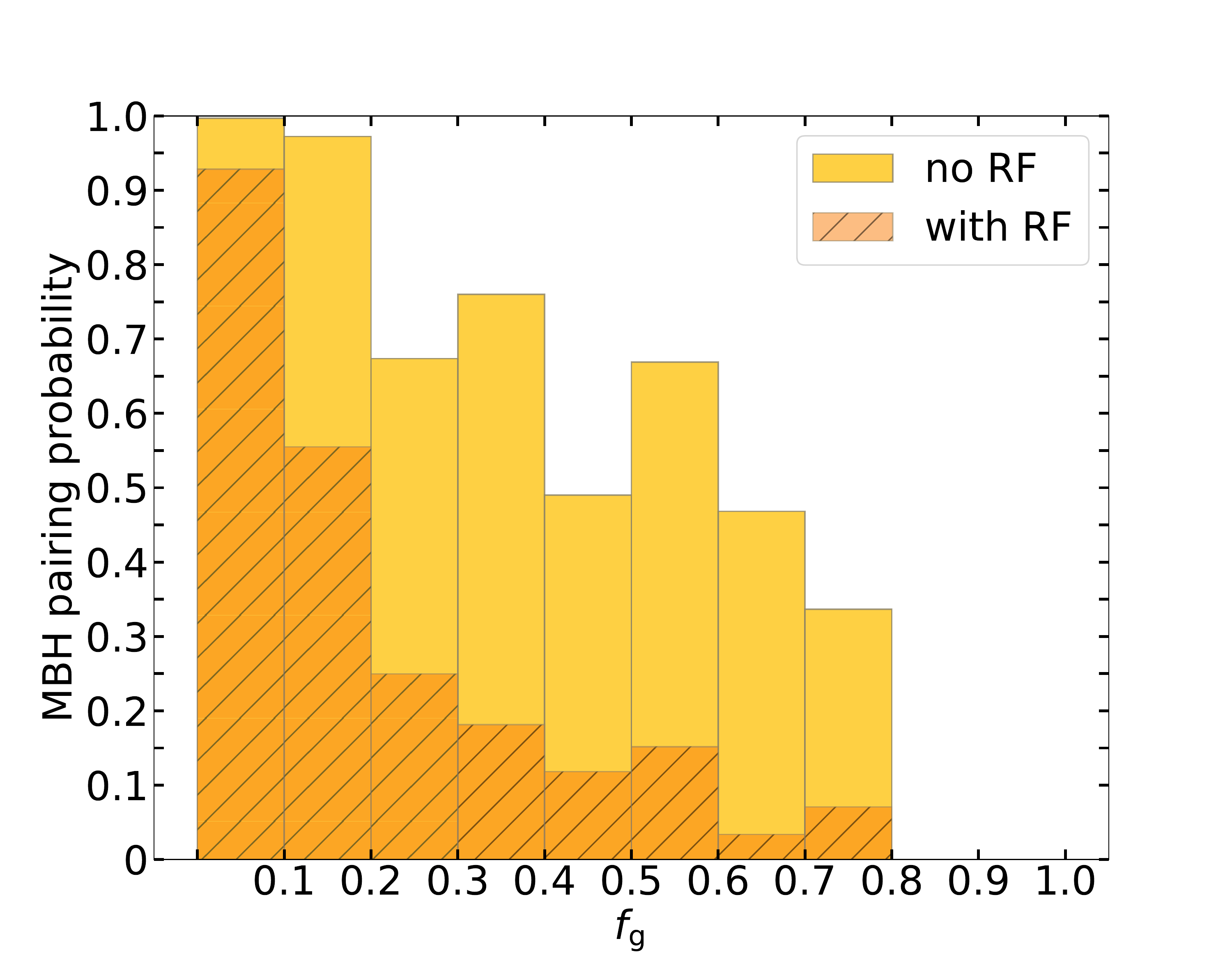}
            \includegraphics[width=0.49\textwidth]{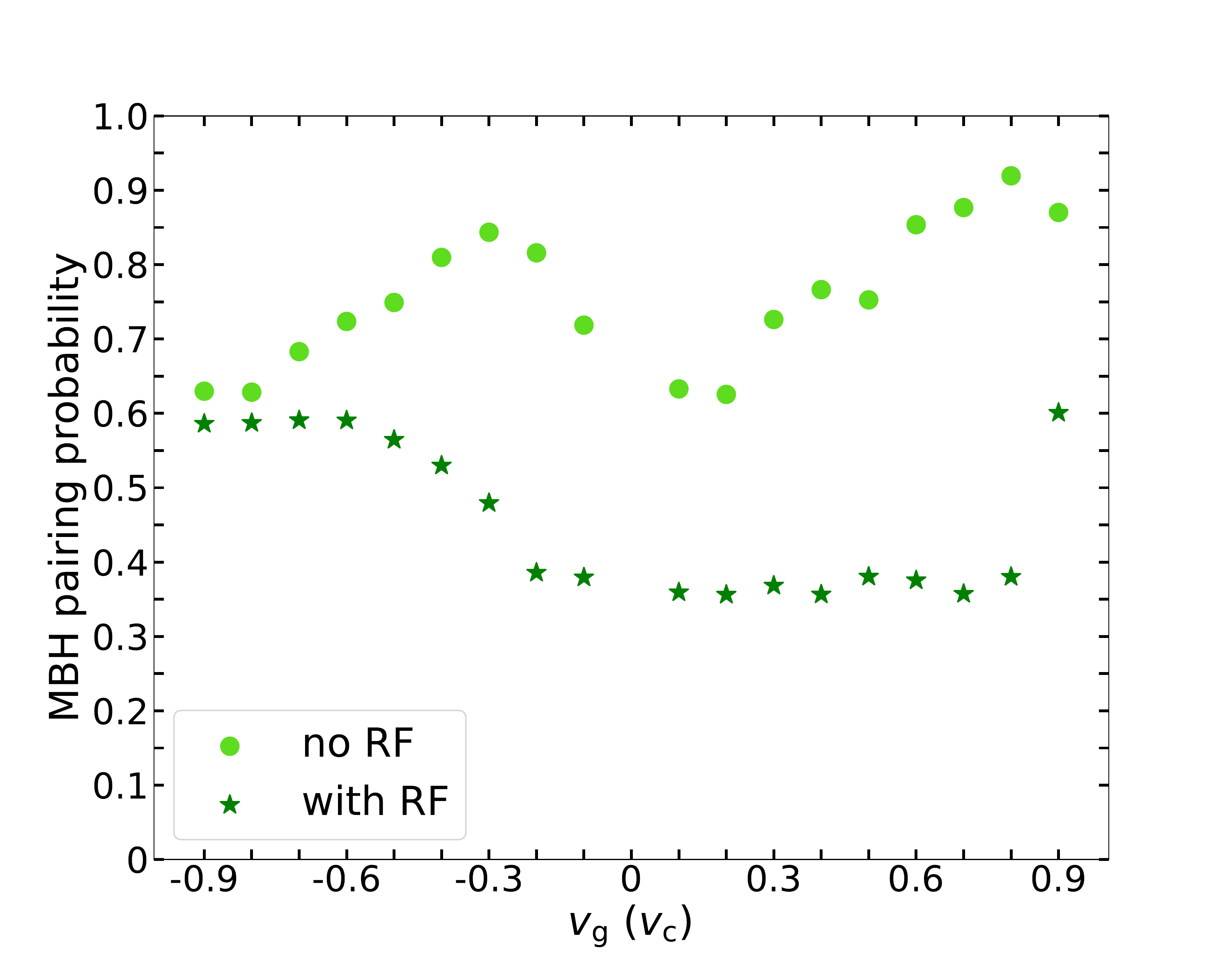}
        \end{tabular}
\caption{MBH pairing probability as a function of the host galaxy and MBH pair properties with and without radiative feedback: \Mtot\ (top left), $q$ (top right), $f_{\rm g}$ (bottom left), and \vg\ (bottom right). We show the dependence on $f_{\rm g}$ as a histogram, since this is a derived, rather than a primary parameter with assumed equidistant values.}
\label{prob_compare}
\end{figure*}

Figure~\ref{prop_l} shows configurations with sMBHs on prograde orbits
with low $e_{\rm i}$, where each data point corresponds to one
simulation. The y-axis of each panel shows the time-averaged azimuthal
component of the DF force, $F_{\phi}$, which dominates the orbital
evolution and includes contributions from the gas
and stellar disk, and stellar bulge. The DF force is in units of
$F_{\rm g,0}= 4\pi\, m_{\rm p}n_{\rm gd0}\,(GM_2/c_{\rm s})^2 =
3.7\times 10^{31}$~dyn, evaluated for $n_{\rm gd0}=100$\,cm$^{-3}$,
$M_2=10^6\,$M$_{\rm \odot}$ and $c_s=10$\,km\,s$^{-1}$. A comparison
of the top and bottom left panels in Figure~\ref{prop_l} shows a
relative dearth of points in simulations with radiative feedback when
$f_{\rm g}\geq0.1$. These configurations are missing because their
\tevol\ becomes longer than a Hubble time. They are the same
population of MBH pairs that contribute to the second peak in the
histograms of Figure~\ref{time_hist} (left panel). Without
radiative feedback their \tevol\ is long and comparable to $\sim
10$\,Gyr, because they experience relatively weak DF force
($|\vec{F}_{\phi}|< 10^{-1} F_{\rm g,0}$). In the presence of
radiative feedback, negative DF prevents their gravitational pairing
within a Hubble time. Our simulations indicate that, in the presence of negative DF, the pairing probability\footnote{Defined as the fraction of simulations in which the sMBH reaches a separation of 1\,pc from the pMBH within a Hubble time.}
%
of MBHs in near circular prograde orbits in galaxies with $f_{\rm g}
\geq 0.1$ is reduced by $91\%$, while in galaxies with $f_{\rm g} <
0.1$ it is only slightly reduced by $1\%$.

The comparison of the right panels of Figure~\ref{prop_l} shows that a
significant fraction of MBH pairs on nearly circular prograde orbits
with $0.2 < v_{\rm g}<0.9$ fail to form bound binaries within a Hubble
time in the presence of radiative feedback. This is particularly true
for systems experiencing a weaker DF force, $|\vec{F}_{\phi}|< 10^{-1}
F_{\rm g,0}$, characteristic of galaxies where gas DF tends to
dominate over stellar DF. This can be understood because the relative
velocities of such sMBHs tend to satisfy ${\cal M}<4$, a necessary
condition for the onset of negative DF.

Figure~\ref{prop_h} is similar to Figure~\ref{prop_l} but shows
sMBHs in prograde orbits with large $e_{\rm i}$ ($e_{\rm i} >
0.8$). The effect of negative DF is now more subtle but still noticeable as a paucity of data points for galaxies with $f_{\rm g}>0.1$ (left panels). 
Collecting all the results from Figs.~\ref{prop_l} and~\ref{prop_h},
we find that negative DF reduces the average pairing probability of MBHs with initially
eccentric orbits by $27\%$, and for those in near circular orbits the probability is
reduced by $50\%$. This difference can be understood by envisioning
the action of  the DF force in each scenario. The velocity of a
prograde sMBH at the apocenter of an eccentric orbit is low relative
to that of the gas disk, which classically results in the DF force in
the direction of motion, acting to speed up the sMBH and circularize
its orbit. In the presence of radiative feedback, however, the DF
force reverses direction, decelerating the sMBH and increasing its
orbital eccentricity. The inspiral time of such eccentric sMBH is
among the shortest in our simulations, since they plunge into the
central parsec instead of going through a lengthy inspiral process,
like nearly circular sMBHs. These systems appear as the deep blue dots
with $|\vec{F}_{\phi}|<10^{-2}$ in the right bottom panel of
Figure~\ref{prop_h}. As a result, the pairing probability of MBHs in
eccentric prograde orbits is not as severely reduced with radiative feedback compared to MBHs in circular prograde orbits.

Overall, we find that, for the full range of galaxy and MBH properties
considered in this work (Table~\ref{tab:params}) and possible orbital
configurations (prograde, retrograde, low/high $e_{\mathrm{i}}$), negative DF reduces the MBH pairing
probability by 46\%. In Figure~\ref{prob_compare} we collect the
results from our entire simulation suite and show this probability as a
function of several key parameters in our model. The top left panel
shows that radiative feedback reduces the average pairing probability of MBH
pairs with total mass in the range of $2\times10^5 M_{\rm \odot} <
M_{\rm bin} < 10^8 M_{\rm \odot}$ from 0.61 to 0.26 (a reduction of 57\%), while the average pairing probability of MBH pairs with masses equal to or larger than
$ 10^8 M_{\rm \odot}$ is nearly unaffected. This happens because
the effect of negative DF, which is more severe for lower mass MBHs
(as indicated by the first criterion in equation~\ref{eq:cond1}), is
compounded with the inefficiency of the DF drag for lower mass objects in general \citep{PB2017}.

The top right panel of Figure~\ref{prob_compare} shows that for MBH
pairs of all masses the pairing probability increases with $q$. It
  nevertheless remains systematically lower by about 35\% in
  simulations that account for the effect of radiative feedback. The
bottom left panel shows that the pairing probability decreases with
the galaxy gas fraction. This trend is present in simulations with and
without radiative feedback. The difference between the two scenarios
is slight in galaxies with $f_g < 0.1$, where the negative DF reduces
the average pairing probability
by $\sim 7\%$. In galaxies with $f_g \geq 0.1$ however, the average
pairing probability is reduced from 0.62 without radiative
feedback to 0.19 with radiative feedback (a reduction of
$\sim 70\%$). This is consistent with the dependence of the DF force
on $f_g$ discussed earlier using the subset of models shown in Figures~\ref{prop_l} and \ref{prop_h}.

The bottom right panel of Figure~\ref{prob_compare} illustrates the
dependence of the pairing probability on \vg. Without
radiative feedback the pairing probability peaks at $v_{\rm
  g}=0.8v_{\rm c}$ and $v_{\rm g}= - 0.3v_{\rm c}$. The peak at
$v_{\rm g}=0.8v_{\rm c}$ is due to sMBHs in circular orbits that
experience efficient gaseous DF. The peak at $v_{\rm g}= - 0.3v_{\rm
  c}$ is due to sMBHs in eccentric orbits whose eccentricity continues
to increase, resulting in them plunging into the central parsec. In
comparison, in simulations with radiative feedback the pairing is most
severely suppressed for prograde MBH pairs in disks with $v_{\rm g} =
0.8 v_{\rm c}$ and  also $v_{\rm g} = - 0.2 v_{\rm c}$, due to the
effect of negative DF. Consequently, the average pairing probability
of MBH pairs in prograde orbits is reduced from 0.78 without
radiative feedback to 0.39 with radiative feedback (a
reduction of 50\%). For MBH pairs in retrograde orbits, the average
  pairing probability is reduced from 0.73 without radiative
  feedback to 0.52 with radiative feedback (a reduction of 28\%).

\section{Potential Impact of Assumptions}
\label{sec:discuss}

We use a semi-analytic model to evaluate the impact of negative DF on
the inspiral time and pairing probability of unequal mass MBH pairs
that evolve under the combined influence of stellar and gaseous DF in
merger remnant galaxies. The power of using the semi-analytic approach is the ability to compute a large number of simulations of MBH orbital decay, over a wide range of galaxy and MBH properties but at the cost of making some simplifying assumptions. We summarize the most important assumptions and their impact below and direct the reader to LBB20 for more details. 
\begin{itemize}
\item The pMBH is fixed at the center of the host galaxy. If the
  motion of the pMBH was instead captured in our simulations, the resulting inspiral times for the modeled MBH pairs would be shorter, particularly in systems with comparable mass MBHs. 

\item The mass of the two MBHs is assumed to remain constant during
  the inspiral, even in galaxies with substantial gas
  fractions. Accretion onto the MBHs during inspiral may change their
  mass ratio and impact the properties of the evolution. While the exact impact depends on the
  details of the accretion onto the MBHs \citep[e.g.,][]{siwek2020},
  an increase in the total mass of the MBH pair will result in a
  shorter inspiral time.  

\item The sMBH is assumed to be completely stripped of its remnant
  stellar cluster during our simulations. This is a plausible outcome
  for our starting radius of $\approx 1$~kpc \citep{KBH2016}. If some
  portion of the stellar cluster survives until late into the
  inspiral, it would lead to more efficient DF and a shorter orbital
  evolution time of the MBH pair \citep{DA2017}. 

\item The orbit of the sMBH is assumed to be co-planar with the galactic gas and stellar disks. It is in nevertheless possible that some fraction of sMBHs evolve on orbits that are inclined relative to the galactic disk. sMBHs on inclined orbits on the one hand experience weaker DF from the gas and stellar disks, an effect that leads to longer inspiral times. On the other hand, perturbations triggered by pericentric passages of the sMBH crossing the disk of the remnant galaxy can trigger the formation of a dense stellar cusp around the sMBH. This effect leads to the increase of stellar mass bound to the sMBH and can shorten the orbital evolution time \citep{W2014}.

\item The gas disk in our model is smooth and devoid of spiral arms or gas clumps. When they are present, interactions between the sMBH and these structures can lead to a random walk of the sMBH, resulting in longer orbital evolution time. In some cases, when the inhomogeneities are large enough, the sMBH may even be ejected out of the galactic disk \citep{T2016}.
\end{itemize}

\section{Conclusions}
\label{sec:concl}
We find that, for a wide range of galaxy and
MBH properties, negative DF reduces the MBH pairing probability by
46\%. In addition, we find that:
\begin{itemize}
\item The effect of negative DF is most pronounced in galaxies with
  significant gas fractions, where gas DF determines orbital evolution
  of the MBH pairs. For example, in galaxies with $f_g \geq 0.1$
  negative DF results in longer MBH inspiral times and reduces the pairing
  probability by $70\%$.
  In contrast, the pairing probability is only slightly reduced in galaxies with $f_g < 0.1$, in which MBH pairs mostly evolve under the influence of stellar DF.

\item Negative DF has a stronger impact on MBHs in prograde orbits (as
  opposed to those in retrograde orbits), since their Mach numbers are
  more likely to have values ${\cal M}< 4$ and fulfill a necessary
  criterion for the onset of negative DF. 
    Similarly, MBH pairs in low $e_{\mathrm{i}}$ orbits are more significantly affected by negative DF than
    those in large $e_{\mathrm{i}}$ orbits.
    This happens because negative DF tends to promote eccentricity growth of already eccentric orbits. The inspiral time of such eccentric MBHs is among the shortest in our simulations, since they plunge into the central parsec instead of going through the lengthy inspiral process.
\item We find that negative DF reduces the pairing probability of
  MBH pairs with total mass $< 10^8 M_{\rm \odot}$ by
  57\%. The effect of negative DF, which is more
  severe at the lower mass end of the MBH spectrum, is compounded with
  the inefficiency of the DF drag for lower mass objects in general. This is of importance because MBH pairs in this mass range are expected to be direct progenitors of merging binaries targeted by the future space-based GW observatory LISA. Specifically, if negative DF operates as described here, the merger rates of MBHBs detectable by LISA may be substantially reduced.
\end{itemize}

Overall, we conclude that negative DF generated by the ionizing
  radiation produced by the inspiralling sMBH is a potentially
  important dynamical effect on the evolution of MBH pairs in
  post-merger galaxies. Future numerical investigations of the
  formation of MBHBs should consider the influence of negative DF,
  in particular in gas-rich galaxies with pair masses $<
  10^8$~M$_{\odot}$.


\acknowledgments

T.B. acknowledges the support by the National Aeronautics and Space
Administration (NASA) under award No. 80NSSC19K0319 and by the
National Science Foundation (NSF) under award No. 1908042. The authors
thank Fabio Antonini for helpful comments.

\bibliographystyle{aasjournal}

\end{document}